\documentclass[amsmath,amssymb,nofootinbib,notitlepage,superscriptaddress,twocolumn]{revtex4-2}

\usepackage{esint,graphicx}

\usepackage{amsmath,amsfonts,amsthm,bm}

\usepackage{stackengine}

\usepackage{rotate}
\usepackage{color}

\DeclareFontFamily{OT1}{pzc}{}
\DeclareFontShape{OT1}{pzc}{m}{it}%
            {<-> s * [1.10] pzcmi7t}{}
\DeclareMathAlphabet{\mathscr}{OT1}{pzc}%
                                {m}{it}

\definecolor{RedWine}{rgb}{0.743,0,0}
\definecolor{RoyalBlue}{rgb}{0.25,.41,.88}
\definecolor{ForestGreen}{rgb}{.13,.54,.13}

\usepackage[backref,breaklinks,colorlinks,citecolor=ForestGreen]{hyperref}

\newcommand{\be}{\begin{equation}}
\newcommand{\ee}{\end{equation}}
\newcommand{\bea}{\begin{eqnarray}}
\newcommand{\eea}{\end{eqnarray}}
\def\ba#1\ea{\begin{align}#1\end{align}}

\begin{document}

\title{Electromagnetic Scoot}

\author{Samuel E. Gralla}
\author{Kunal Lobo}
\affiliation{Department of Physics, University of Arizona, Tucson, Arizona 85721, USA}

\begin{abstract}
Recent work on scattering of massive bodies in general relativity has revealed that the mechanical center of mass of the system (or, more precisely, its relativistic \textit{mass moment}) undergoes a shift during the scattering process.  We show that the same phenomenon occurs in classical scattering of charged particles in flat spacetime and study the effect in detail.  Working to leading order in the interaction, we derive formulas for the initial and final values of the mechanical and electromagnetic energy, momentum, angular momentum, and mass moment.  We demonstrate that the change in mechanical mass moment is balanced by an opposite change in the mass moment stored in the electromagnetic field.  This is a non-radiative exchange between particles and field, analogous to exchange of kinetic and potential energy.  A simple mechanical analogy is a person scooting forward on the floor, who exchanges mass moment with the floor.  We therefore say that electromagnetic scattering results in an electromagnetic scoot.
\end{abstract}

\maketitle

\section{Introduction}

Scattering experiments, whether real or fictitious, offer a simple way to gain understanding of the physical implications of a theory.  The small-deflection limit provides a further simplified testing ground where precise analytical results are usually possible.  The study of small-angle scattering of comparable-mass particles in general relativity began in the 1980's with the derivation of the first-order \cite{Portilla:1980uz} and second-order \cite{Westpfahl:1985tsl} deflection angle, and has recently seen a resurgence of interest due to connections with quantum scattering methods and the dynamics of bound systems.  The new computational firepower thrown at this problem has resulted in spectacular progress, with the latest results now probing the \textit{fourth} order in the small-angle approximation (\cite{Bern:2021dqo,Dlapa:2021npj,Bern:2021yeh,Dlapa:2021vgp} and references therein).  

Inspired by this rich, interconnected set of results, we set about to understand gravitational scattering with a new approach using self-force methods \cite{Gralla:2021qaf}.  We managed to reproduce the second-order results from the 1980s, but discovered, to our surprise, an overlooked feature of the problem that appears even at the \textit{first} order beyond straight line motion.  In addition to computing the energy, momentum, and angular momentum of the particles, we considered the last, overlooked conserved quantity: the mass moment.  For a system of point particles, the mass moment is defined by
\begin{align}\label{Nmechintro}
    \bm{N}_{\rm mech} = \sum_I E_I \bm{r}_I - t \sum_I \bm{p}_I,
\end{align}
where $\bm{r}_I$, $E_I$ and $\bm{p}_I$ are the position, energy and momentum (respectively) of the particles labeled by $I$.\footnote{We set the speed of light to unity ($c=1$) and regard relativistic mass and energy as equivalent.  If we had not made this choice, we would divide the first term of \eqref{Nmechintro} by $c^2$, ensuring that mass moment has units of mass times length.}  We include the subscript ``mech'' to emphasize that any field contributions have not been included in this formula. 

The mass moment is the position-weighted energy of the system minus its total momentum times time, and its conservation reflects the uniform motion of the center of energy.  It is numerically equal to the total energy times the center of energy at time $t=0$, and it  thereby tracks the position of the center of energy at a fiducial time.  Although the total value can always be set to zero by a translation, the mass moment is additive (unlike the center of energy) and can therefore be budgeted like the energy, momentum, and angular momentum.  That is, we can ask about \textit{exchange} of mass moment between different degrees of freedom, or \textit{radiation} of mass moment away to infinity.  From a relativistic point of view, mass moment is inseparable from the angular momentum, since the two mix under boosts and only together form a relativistically invariant object (e.g., \cite{MTW}).

In the scattering of two masses $m_1$ and $m_2$, the important mass scales are the initial total energy $E_0$ and the relativistic reduced mass $\mu$,
\begin{align}\label{E0}
    E_0 = \sqrt{m_1^2 + m_2^2 + 2\gamma m_1 m_2}, \qquad \mu = \frac{m_1 m_2}{E_0},
\end{align}
where $\gamma$ is the initial relative Lorentz factor.  In the small-deflection limit, the center of energy-momentum (CEM) frame scattering angle is proportional to \cite{Westpfahl:1985tsl}
\begin{align}
\chi = \frac{G E_0}{bv^2} \ll 1,
\end{align}
where $G$ is Newton's constant, $b$ is the impact parameter, and $v$ is the initial relative velocity.  In our study of small-angle gravitational scattering through second order in $\chi$ \cite{Gralla:2021qaf}, we found that the mechanical mass moment changes during the scattering process.  At leading order in $\chi$, the change is
\begin{align}
    \Delta \bm{N}_{\rm mech} = 2 \mu b \chi \gamma(1-3v^2) \log \frac{m_2 + \gamma m_1}{m_1 + \gamma m_2} \bm{\hat{z}},
\end{align}
where $\bm{\hat{z}}$ is a unit vector pointing  from particle 1 to particle 2 at early times. This form makes clear that the effect disappears in the Newtonian limit $v \to 0$, as it must.\footnote{The appropriate Newtonian limit is $v \to 0$ at fixed $b$ and $\chi$, which preserves the condition $\chi \ll 1$ (small-angle deflection) assumed in our calculation.  If one instead fixes $b$ and $E_0$, the limit describes a bound system for which our scattering results are invalid.}   However, plugging in for $\chi$ shows that the change in mass moment is in fact \textit{independent of the impact parameter},
\begin{align}\label{gravscoot}
    \Delta \bm{N}_{\rm mech} = 2 \gamma\frac{1-3v^2}{v^2} G m_1 m_2 \log \frac{m_2 + \gamma m_1}{m_1 + \gamma m_2} \bm{\hat{z}}.
\end{align}
This suggests that the effect is not tied to the details of the small-angle scattering encounter and will exist in similar form for large-angle scattering with $\chi \gtrsim 1$.

These results surprised and puzzled us for a number of reasons, not least because we believed we were working in the CEM frame, where the total mass moment should be zero.  We suspected that contributions from the gravitational field need to be included, but ran into obstacles related to the fact that field energy is fundamentally gauge-dependent in general relativity.  It is also not entirely clear that spacetime can be treated as flat for the purposes of computing conserved quantities at early and late times, since these involve $1/t$ corrections that are formally the same order as gravitational field perturbations to the metric.

Work is underway to settle these issues in general relativity \cite{gravscoot}.  However, in the meantime, we can consider a simpler, electromagnetic analog where the problems do not arise.  This problem is surely even older than the gravitational one, and many aspects of the calculation have undoubtedly been performed before (not least in the recent, mammoth exploration through third perturbative order \cite{Saketh:2021sri}).  However, we are unaware of any results on the mass moment, or even any mention of this quantity, in past work on  electromagnetic scattering.

Consider, then, the small-angle scattering of classical charged particles.  The leading deflection angle is proportional to \cite{Westpfahl:1985tsl}
\begin{align}
    \chi_{\rm EM} = \frac{q_1 q_2}{\mu b v^2} \ll 1,\label{chiEM}
\end{align}
where $q_1$ and $q_2$ are the particle charges. 
(There is no explicit coupling constant analogous to $G$ because we work in Gaussian units.)
Computing at first order in $\chi_{\rm EM}$, we find a precisely analogous change in mass moment [Eq.~\eqref{DeltaNmech} below],
\begin{align}
    \Delta \bm{N}_{\rm mech} & = -2 \mu b \gamma^{-2} \chi_{\rm EM} \log\frac{m_2 + \gamma m_1}{m_1 + \gamma m_2} \bm{\hat{z}} \label{NEM1} \\
    & = -\frac{2 q_1q_2}{\gamma^2 v^2}\log\frac{m_2 + \gamma m_1}{m_1 + \gamma m_2} \bm{\hat{z}}. \label{NEM2}
\end{align}
Eq.~\eqref{NEM1} demonstrates the relationship to our perturbative calculation, while Eq.~\eqref{NEM2} shows that the change in mechanical mass moment is again independent of the impact parameter.

In the electromagnetic setting we can be confident, based on general theorems, that this change is in fact compensated by an equal and opposite change in the electromagnetic contribution to the mass moment.  The EM field mass moment is given by
\begin{align}\label{NEM}
    \bm{N}_{\rm EM} = \int \mathcal{E} \bm{x} d^3 x - t \int \bm{S} d^3x,
\end{align}
where $\mathcal{E}$ and $\bm{S}$ are the electromagnetic field energy and momentum densities, respectively.  We argue that only the cross-term contributions (proportional to $q_1 q_2$) should be included in the point particle limit and explicitly evaluate these integrals at early and late times.   We find that, indeed, the electromagnetic contribution exactly balances the mechanical one,
\begin{align}
    \Delta \bm{N}_{\rm EM} = - \Delta \bm{N}_{\rm mech}.
    %\bm{N} = \bm{N}_{\rm mech} + \bm{N}_{\rm EM} = 0.
\end{align}
That is, there is no change in total mass moment, only an \textit{exchange} between mechanical and electromagnetic degrees of freedom.  The exchange is permanent: an electromagnetic scoot.  

The electromagnetic problem thus provides a neat and tidy story that can be fully understood.  The details and outcome of this calculation  give insight into electromagnetic phenomena and lessons for seeking analogous understanding in the gravitational case.  We discuss these points in detail at the conclusion of the manuscript.

While this paper was inspired by gravitational phenomena and is aimed substantially at researchers working in this area, we feel that the electromagnetic results are of interest in their own right.  As such, we have endeavored to make the paper accessible to aficionados of electromagnetism who are not necessarily steeped in relativistic notation.  We have therefore chosen to use vector notation throughout, eschewing the tensors that are standard in gravitational physics.  However, we have retained the use of Gaussian units with the speed of light set equal to one, so that results can be easily compared with electromagnetic calculations in the high-energy and gravitational physics literature.  Readers unfamiliar with these units can always restore constants like $4\pi \epsilon_0$ and $c$ via dimensional analysis.

This paper is organized as follows.  In Sec.~\ref{sec:scattering} we set up the problem and derive the particle trajectories and electromagnetic fields through first order in the interaction.  In Sec.~\ref{sec:conserved} we calculate all conserved quantities at leading order at early and late times ($t \to \pm \infty$) and analyze the implications.  We pay particular attention to the mass moment and also discuss the center of energy.  We conclude with some discussion of the scoot phenomenon.  An appendix describes the evaluation of certain integrals that arise in the analysis.

\section{small-angle scattering of relativistic charged particles}\label{sec:scattering}

We will consider a scattering encounter between two classical charged particles $1$ and $2$ in the approximation of small deflection.  To leading order, the particles move in straight lines, and the description is simplest in the frame where one particle is at rest.  We will take particle $2$ to be at rest at the origin and denote this frame with a prime.  Choosing the motion to be in the $z'$ direction and the transverse separation to be in the $x'$ direction, the leading-order trajectories are simply 
\begin{align}
\bm{r}_1' & = (b,0,v t') \label{r1p} \\
\bm{r}_2' & = (0,0,0), \label{r2p}
\end{align}
where $b$ and $v$ are constants interpreted as the impact parameter and relative velocity, respectively.  Without loss of generality we assume that $v$ is positive,
\begin{align}
    v>0,
\end{align}
so that particle 1 moves in the $+z'$ direction.

We determine the corrected motion by integrating the Lorentz force law using the electric and magnetic fields produced by the background trajectories.  There are no magnetic forces since $\dot{\bm{r}}_2'=0$ and $\bm{B}_2'=0$ to zeroth order, and the Lorentz force law becomes
\begin{align}
m_1 \ddot{\bm{r}}_1' & = \frac{q_1}{\gamma} \left( \bm{E}_2' - v^2 (\hat{\bm{z}}'\cdot \bm{E}_2') \hat{\bm{z}}'\right) \label{a1p}\\
m_2 \ddot{\bm{r}}_2' & = q_2 \bm{E}_1',\label{a2p}
\end{align}
where $\gamma=(1-v^2)^{-1/2}$ is the relative Lorentz factor.  The electric fields produced at leading order are 
\begin{align}
\bm{E}_1' & = q_1 \gamma\frac{(x'-b)\bm{\hat{x}}' + y'\bm{\hat{y}'} + (z'-vt')\bm{\hat{z}}'}{[(x'-b)^2+y'^2+\gamma^2(z'- v t')^2]^{3/2}} \label{E1p} \\
\bm{E}_2' & = q_2\frac{x'\bm{\hat{x}}' + y' \bm{\hat{y}}'+ z'\bm{\hat{z}}'}{[x'^2+y'^2+z'^2]^{3/2}}.\label{E2p}
\end{align}
The corrected motion is determined by integrating the right-hand-sides of Eqs.~\eqref{a1p} and \eqref{a2p} using Eqs.~\eqref{E1p} and \eqref{E2p}.  We choose the integration constants so that the perturbed velocity of each particle vanishes in the distant past (making $v$ interpreted as the initial relative velocity), and so that the particles reach $z'=0$ at $t'=0$.  We find
\begin{align}
x_1' & = b + \frac{q_1 q_2}{b m_1 \gamma v^2}\left( vt' + \sqrt{b^2+v^2 t'^2} \right) \label{x1p} \\
z_1' & = v t' - \frac{q_1 q_2}{m_1 \gamma^3v^2}\textrm{arctanh}\frac{vt'}{\sqrt{v^2t'^2+b^2}} \label{z1p}\\
x_2' & = -\frac{q_1 q_2}{b m_2  v^2}\left( vt' + \sqrt{v^2 t'^2+b^2 \gamma^{-2}} \right)\label{x2p} \\
z_2' & = \frac{q_1 q_2}{m_2 \gamma^2v^2}\textrm{arctanh}\frac{vt'}{\sqrt{v^2t'^2+b^2\gamma^{-2}}}.\label{z2p}
\end{align}
These equations provide the corrected particle trajectories.  The second particle is no longer at rest, but its velocity is asymptotically zero at early times.  We may thus interpret the primed frame as the ``initial rest frame'' of particle 2.  Note, however, that the position of particle 2 in fact diverges logarithmically at early (and late) times.  This is an unavoidable consequence of the long-range nature of the Coulomb force: an inverse-square force integrates up to a logarithmically divergent position.

\subsection{Transformation to CEM frame}

While the primed frame was convenient for finding the trajectories, it is conceptually less useful since it makes an explicit preference for one particle over the other, when no such preference exists in the problem.  A more natural choice is the center of energy-momentum (CEM) frame, defined as the frame with no momentum $\bm{p}$ or mass moment $\bm{N}$,
\begin{align}\label{CEM-def}
\bm{p} = \bm{N} = 0.
\end{align}
In our relativistic scattering problem, these quantities receive contributions from both the particles and the electromagnetic field (see Sec.~\ref{sec:conserved}).  At leading order (neglecting the interaction), the particles move in straight lines [Eqs.~\eqref{r1p} and \eqref{r2p}] and there is no contribution from the electromagnetic field.  In this case the appropriate transformation consists of a Lorentz boost in the $z$ direction and a translation in the $x$ direction,
\begin{align}
t & = \frac{m_2+\gamma m_1}{E_0}t' - \frac{\gamma m_1 v}{E_0}z' \label{tCM} \\
z & = \frac{m_2+\gamma m_1}{E_0}z' - \frac{\gamma m_1 v}{E_0}t' \label{zCM} \\
x & = x' - b\frac{m_1(\gamma m_2  + m_1)}{E_0^2}, \label{xCM}
\end{align}
where $E_0$ was introduced in Eq.~\eqref{E0} above.

Although this transformation was designed for the leading order motion, it turns out that no modification is necessary for the first perturbative correction.  That is, we will see that Eqs.~\eqref{tCM}-\eqref{xCM} still take us to the CEM frame as defined by \eqref{CEM-def}.  However, and quite surprisingly, we find that it is essential to take into account the contribution from the electromagnetic field, \textit{even at early and late times, when the particles are infinitely separated}.  This fact enables the electromagnetic scoot: a permanent exchange of mass moment between particle and field.

\subsection{CEM-frame particle trajectories}

The path of particle $1$ in the CEM frame is given by plugging Eqs.~\eqref{x1p} and \eqref{z1p} into Eqs.~\eqref{tCM}--\eqref{xCM} and solving the resulting set of equations for $x_1(t)$ and $z_1(t)$, dropping terms nonlinear in $q_1 q_2$; the same procedure for particle $2$ gives $x_2(t)$ and $z_2(t)$.  The results are

\begin{align}
x_1 &=  b_1 + \frac{q_1 q_2}{b m_1 \gamma v^2}\left( v t_1 + \sqrt{b^2+v^2 t_1^2} \right)  \label{x1} \\
z_1 &=v_1 \left( t -  \frac{q_1 q_2 E_0}{m_1 m_2 \gamma^3 v^3}\textrm{arctanh} \frac{v t_1}{\sqrt{b^2 + v^2 t_1^2}} \right) \label{z1} \\
x_2 &= b_2 - \frac{q_1 q_2}{b m_2 \gamma v^2}\left( v t_2 + \sqrt{b^2+v^2 t_2^2} \right) \label{x2} \\
z_2 &= v_2\left( t - \frac{q_1q_2 E_0}{m_1 m_2 \gamma^3 v^3}\textrm{arctanh}\frac{v t_2}{\sqrt{b^2 + v^2 t_2^2}}\right), \label{z2}
\end{align}

where we define
\begin{align}
t_1 & = \frac{\gamma E_0}{m_1 + \gamma m_2}t, \qquad \! \ \ \ \ t_2= \frac{\gamma E_0}{m_2 + \gamma m_1}t \label{tdefs} \\
b_1 & =  \frac{m_2(m_2+\gamma m_1)}{E_0^2}b, \quad
b_2  =  -\frac{m_1(m_1+\gamma m_2)}{E_0^2}b \label{bdefs} \\
v_1 & = \frac{\gamma m_2}{m_1 + \gamma m_2}v, \qquad \ \ \ 
v_2  = -\frac{\gamma m_1}{m_2 +\gamma m_1}v. \label{vdefs}
\end{align}
Notice that relabeling the particles ($m_1 \leftrightarrow m_2$ and $q_1 \leftrightarrow q_2$) is equivalent to reversing the sign of $x$ and $z$.  In other words, the configuration is invariant under swapping the particles and also rotating by $180^\circ$ within the plane of their motion.  This  symmetry is a special property of the CEM frame.  It was taken as the \textit{definition} of the CEM frame in our recent work \cite{Gralla:2021qaf}.

\subsection{CEM-frame electromagnetic field}

The leading electromagetic field is determined by the leading, straight-line motion of the charges.  Since the sources have constant velocity, their fields are just the boosted Coulomb field, given for $I=1,2$ by
\begin{align}
\mathbf{E}_I & = \frac{q_I\gamma_I}{R_I^3}\left[ (x-b_I)\bm{\hat{x}} + y \bm{\hat{y}} + (z-v_I t)\bm{\hat{z}} \right] \label{EI} \\ %(x-b_I, y, z-v_It) \\
\mathbf{B}_I & = \frac{-q_I\gamma_I v_I}{R_I^3}\left[ y \bm{\hat{x}} -(x-b_I) \bm{\hat{y}} \right], \label{BI} %(y, -(x-b_I), 0)
\end{align}
where $\gamma_I=(1-v_I^2)^{-1/2}$ and the boosted distance function is 
\begin{align}
    R_I & = \sqrt{(x-b_I)^2+y^2+\gamma_I^2(z-v_It)^2}.
\end{align}
Notice that the electric and magnetic fields are invariant under the operations $m_1 \leftrightarrow m_2$, $q_1 \leftrightarrow q_2$,  $x \to -x$, $z\to -z$, a special property of the CEM frame [see discussion below Eq.~\eqref{vdefs}].

This completes the first-corrected description of the problem in the CEM frame: the particle trajectories are given in Eqs.~\eqref{x1}--\eqref{z2}, while the electric and magnetic fields are given in Eqs.~\eqref{EI} and \eqref{BI}. 

\section{Analysis of conserved quantities}\label{sec:conserved}

We will now discuss the behavior of the four conserved quantities: energy, momentum, angular momentum, and mass moment.  For clarity, let us first imagine the situation where the particles are modeled by smooth, extended bodies.  The budget for the system involves mechanical contributions from bodies $1$ and $2$ as well as the contribution from the electromagnetic field,
\begin{align}
E & = E_1 + E_2 + E_F \\
\bm{p} & = \bm{p}_1 + \bm{p}_2 + \bm{p}_{F} \\
\bm{L} & = \bm{L}_1 + \bm{L}_2 + \bm{L}_{F} \\
\bm{N} & = \bm{N}_1 + \bm{N}_2 + \bm{N}_{F}.
\end{align}
The form of the body contributions will depend on the particular model for the bodies, but the electromagnetic contribution is always given by
\begin{align}
E_F & = \frac{1}{8\pi} \int \left( \bm{E}^2+\bm{B}^2 \right)  d^3 x \label{EF} \\
  \bm{p}_F & = \frac{1}{4\pi}\int \left( \bm{E} \times \bm{B} \right) d^3x \label{pF} \\
  \bm{L}_F & = \frac{1}{4 \pi}\int \bm{x} \times (\bm{E} \times \bm{B}) \ \! d^3 x\\
    \bm{N}_F & = \frac{1}{8\pi} \int \left( \bm{E}^2+\bm{B}^2 \right) \bm{x} \ \! d^3 x - \bm{p}_F t. \label{NF}
\end{align}

Now let us consider the point particle limit.  The particle conserved quantities take their standard relativistic forms,
\begin{align}
E_I & = \gamma_I m_I \label{Ep} \\
\bm{p}_I & = \gamma_I m_I \dot{\bm{r}}_I \label{pp} \\
\bm{L}_I & = \bm{r}_I \times \bm{p}_I \label{Lp} \\
\bm{N}_I & = \gamma_I m_I \bm{r}_I - \bm{p}_I t,\label{Np}
\end{align}
where $I=1,2$ labels the particles.  In these expressions the full Lorentz factors $\gamma_I=(1-\dot{\bm{r}}_I^2)^{-1/2}$ must be used, as opposed to the background Lorentz factors appearing in Eqs.~\eqref{EI} and \eqref{BI}.  The point particle limit is a significant simplification, since we now have definite expressions for the conserved quantities.

The point particle limit will also help with the field integrals, but a naive application brings trouble.  Whereas Eqs.~\eqref{EF}-\eqref{NF} are perfectly well-defined for extended bodies, they are divergent for point particles.  The electric and magnetic fields grow like inverse distance squared as one approaches the particles, so that the densities of the conserved quantities grow like inverse distance to the fourth power.  These singularities are not integrable, and all four conserved quantities are divergent.  This issue is generally known as the  ``electron self-energy problem,'' although the problem is with the naive point particle limit, not with electrons.

To see how to proceed, let us consider the example of the energy,
\begin{align}
    E_{F} = \frac{1}{8\pi} \int \left[(\bm{E}_1 + \bm{E}_2)^2 + (\bm{B}_1 + \bm{B}_2)^2 \right] d^3 x.
\end{align}
This integral naturally splits into three contributions $E_f = E_{F1}+E_{F2}+E_{F\times}$ given by
\begin{align}
    E_{F1} & = \frac{1}{8\pi} \int (\bm{E}_1^2 + \bm{B}_1^2)  d^3 x \\
    E_{F2} & = \frac{1}{8\pi} \int (\bm{E}_2^2 + \bm{B}_2^2)  d^3 x \\
    E_{F\times} & = \frac{1}{4\pi} \int (\bm{E}_1\cdot \bm{E}_2 + \bm{B}_1 \cdot \bm{B}_2) d^3 x.\label{Ecross}
\end{align}
The integrals for $E_{F1}$ and $E_{F2}$ are infinite on account of the inverse-square divergence of the electric and magnetic fields at the positions of the particles.  However, there are several different ways to see that we can, and in fact \textit{must}, drop these terms from the calculation.

The simplest reason is that $E_{F1}$ and $E_{F2}$ are proportional to $q_1^2$ and $q_2^2$, respectively, whereas the correction we consider is proportional to $q_1 q_2$ [see Eqs.~\eqref{x1}--\eqref{z2}].  We know that there is in fact no energy exchange between particle and field at order $q_1^2$ or $q_2^2$ (the self-force effects start at higher order \cite{Saketh:2021sri}), so it is consistent to drop these terms from the energy budget.  Including them properly would actually be quite subtle, since they contribute to the particle masses $m_1$ and $m_2$.\footnote{In the classic derivation of the self-force (e.g. \cite{Poisson:1999tv}), the infinite self-energy is combined with a negatively infinite ``bare mass'', with the sum representing the finite particle mass $m$.  In a rigorous derivation with extended bodies \cite{Gralla:2009md}, one finds an analogous \textit{finite} mass renormalization, with the observable mass \textit{proven} to be a sum of material and field contributions, each of which is individually finite.  This decomposition occurs even in the derivation of the Lorentz force law, irrespective of self-force corrections.}  It is also useful to consider the slow motion limit, where one has the well known conservation of total (kinetic plus potential) energy.  In this limit the cross-term integral evaluates precisely to the usual interaction energy $U=q_1 q_2/(|\bm{r}_1-\bm{r}_2|)$ [see Eq.~\eqref{EFxtilde} below].  That is, reproducing the usual conservation of total energy requires dropping the self-energy terms.

For all these reasons, the correct budget for the conserved quantities in our problem is
\begin{align}
E & = E_1 + E_2 + E_{F\times} \label{Etot} \\
\bm{p} & = \bm{p}_1 + \bm{p}_2 + \bm{p}_{F\times} \label{ptot} \\
\bm{L} & = \bm{L}_1 + \bm{L}_2 + \bm{L}_{F\times} \label{Ltot} \\
\bm{N} & = \bm{N}_1 + \bm{N}_2 + \bm{N}_{F\times},\label{Ntot}
\end{align}
where the particle contributions are given by Eqs.~\eqref{Ep}-\eqref{Np}, while the cross-term field contributions are given by
\begin{align}
E_{F\times} & = \frac{1}{8\pi} \int \mathcal{E}_\times d^3 x \label{EFx} \\
  \bm{p}_{F\times} & = \frac{1}{4 \pi}\int \bm{S}_\times d^3x \label{pFx} \\
  \bm{L}_{F\times} & = \frac{1}{4 \pi}\int \bm{x} \times \bm{S}_\times \ \! d^3 x \label{LFx}\\
    \bm{N}_{F\times} & = \frac{1}{8\pi} \int \mathcal{E}_\times \bm{x} \ \! d^3 x - \bm{p}_{F\times} t. \label{NFx}
\end{align}
Here we have introduced the cross-term energy and momentum densities as
\begin{align}
        \mathcal{E}_\times & = \frac{1}{4\pi} \left( \bm{E}_1 \cdot \bm{E}_2 + \bm{B}_1\cdot \bm{B}_2 \right) \label{ecross} \\
        \bm{S}_\times & = \frac{1}{4\pi}\left( \bm{E}_1 \times \bm{B}_2 + \bm{E}_2 \times \bm{B}_1 \right). \label{Scross}
\end{align}
All terms in the conserved quantity budgets \eqref{Etot}-\eqref{Ntot} are now fully specified and mathematically well-defined.

\subsection{Initial and final values: mechanical contribution}

In order to understand the exchange of conserved quantities between particles and field, we will consider those quantities at early and late times in the scattering problem. Expanding the trajectories at $t \to \pm \infty$, we find
\begin{align}
x_I &= b_I + \Theta(t) \frac{2 q_1 q_2}{\mu b \gamma v^2}v_It + O(1/t^2) \\
z_I &= v_I t \mp v_I \frac{q_1 q_2}{\mu \gamma^3 v^3}\log\frac{2 v |t_I|}{b} + O(1/t^2)
\end{align}
Here and below, the upper sign corresponds to late times, while the lower sign corresponds to early times.  Notice that the $z$ position of both particles is logarithmically divergent at early and late times.  As discussed below Eq.~\eqref{z2p}, the divergence originates from the inverse-square nature of the electromagnetic force.  The presence of the Heaviside function $\Theta(t)$ in the $x$ position shows the scattering of the particles by a small angle $\delta$: 
\begin{align}
    \delta = \lim_{t \to \infty} \left( \frac{x_I- b_I}{z_I}\right)  = \frac{2 q_1 q_2}{\mu b \gamma v^2}.
\end{align}
This result is well known (e.g., \cite{Saketh:2021sri}).

Calculating the conserved quantities from Eqs.~\eqref{Ep}-\eqref{Np}, we have
\begin{align}
E_1 & = \frac{m_1+\gamma m_2}{E_0}\left( m_1 - \frac{m_2}{E_0}\frac{q_1 q_2}{\gamma v |t|}\right) +O(t^{-2})\\
\bm{p}_1 & = \left(\mu \gamma v - \frac{q_1q_2}{\gamma^2 v^2}\frac{(m_1 + \gamma m_2)^2}{E_0^2 |t|}\right)\bm{\hat{z}} \nonumber \\
& \qquad + \Theta(t) \frac{2 q_1 q_2}{bv} \bm{\hat{x}} + O(t^{-2}) \\
\bm{L}_1 & = - \mu b \gamma v \frac{m_2(m_2+\gamma m_1)}{E_0^2} \bm{\hat{y}} + O(t^{-1})  \\
\bm{N}_1 & = \mp \frac{q_1 q_2}{ \gamma^2 v^2}\left(\log  \frac{2 \gamma v E_0 |t|}{(m_1 + \gamma m_2)b}-1 \right) \bm{\hat{z}} \nonumber \\ & \quad + \gamma_1 m_1 b_1 \bm{\hat{x}} +O(t^{-1})
\end{align}

The values for particle 2 are given by exchanging $1 \leftrightarrow 2$ and sending $x \to -x$ and $z \to -z$ [see discussion below Eq.~\eqref{vdefs}].  The energy, momentum, and angular momentum have well-defined initial and final values (good limits as $t \to -\infty$ and $t \to +\infty$, respectively), but the mass moment inherits the logarithmic divergence of the position.  If we instead consider the total mechanical contribution to the conserved quantities, we have 
\begin{align}
E_1 + E_2 & = E_0 - \frac{q_1 q_2}{ v |t|} \frac{m_1^2+m_2^2+2m_1m_2/\gamma}{E_0^2} \label{E12} \\
\bm{p}_1 + \bm{p}_2 & = -\frac{q_1q_2}{|t|}\frac{m_2^2-m_1^2}{E_0^2}\bm{\hat{z}}+O(t^{-2}) \label{p12} \\
\bm{L}_1 + \bm{L}_2 & = - \mu b \gamma v \bm{\hat{y}} + O(t^{-1}) \label{L12} \\
\bm{N}_1 + \bm{N}_2 & = \mp \frac{q_1q_2}{\gamma^2 v^2}\log\frac{m_2 + \gamma m_1}{m_1 + \gamma m_2} \bm{\hat{z}} + O(t^{-1}). \label{N12}
\end{align}
The total mechanical mass moment has well-defined initial and final  values (limits as $t\to \pm\infty$), given by the lower sign and upper sign (respectively) in Eq.~\eqref{N12}.  These values are different, meaning there is a permanent change in mechanical mass moment.  That is, if $\Delta$ represents final minus initial, and ``mech'' refers to the total contribution from the particles, we have 
\begin{align}
    \Delta E_{\rm mech} = 0, \quad \Delta \bm{p}_{\rm mech} = 0, \quad \Delta \bm{L}_{\rm mech} = 0,
\end{align}
but 
\begin{align}
    \Delta \bm{N}_{\rm mech} = -\frac{2 q_1q_2}{\gamma^2 v^2}\log\frac{m_2 + \gamma m_1}{m_1 + \gamma m_2} \bm{\hat{z}}. \label{DeltaNmech}
\end{align}
Since the total mass moment is conserved, there must be an opposing change in the electromagnetic field mass moment.  We will now directly compute the field contributions to the conserved quantities in order to see the exchange explicitly.

\subsection{Initial and final values: field contribution}\label{sec:field-early}

We wish to compute the initial and final values of the field contributions to the conserved quantities.  At early and late times, the particles are widely separated compared to their impact parameter, and one expects the problem to become effectively one-dimensional.  This intuition is confirmed by dimensional analysis, as follows.  At a given time $t$, the problem contains two length scales, $b$ and $v t$.  The cross-term field contributions can depend only on the ratio after a relevant dimensionful combination has been factored out,
\begin{align}
    E_{F\times} & = \frac{q_1 q_2}{v t} f_E\left(\frac{b}{vt}, m_1, m_2, v\right) \label{EFfunc}\\
    \bm{p}_{F\times} & = \frac{q_1 q_2}{v t} f_p\left(\frac{b}{vt}, m_1, m_2, v\right) \label{pFfunc}\\
    \bm{L}_{F\times} & = q_1 q_2 f_L\left(\frac{b}{vt}, m_1, m_2, v\right) \label{LFfunc} \\
    \bm{N}_{F\times} & = q_1 q_2 f_N\left(\frac{b}{vt}, m_1, m_2, v\right). \label{NFfunc}
\end{align}
The functions $f$ are just placeholders indicating functional dependence.  These equations may also be derived mathematically by making the change of variables $\bm{x}'=\bm{x}/(vt)$ in the cross-term integrals \eqref{EFx}-\eqref{NFx} and noting that $b_I/(v_I t)$ is independent of $I$,
\begin{align}
    \frac{b_I}{v_I t} = \frac{b}{vt}\frac{(m_1+ \gamma m_2)(m_2+ \gamma m_1)}{E_0^2 \gamma}.
\end{align}
Eqs.~\eqref{EFfunc}-\eqref{NFfunc} show that the leading behavior at large $|t|$ may be computed using the limit $b \to 0$ at fixed $t$.  This confirms the intuition that the problem is one-dimensional at early (and late) times and allows us to use the $b=0$ versions of the electric and magnetic fields to compute the cross-term field contributions.  These integrals are evaluated in Appendix \ref{sec:field-integrals}.  Noting our convention $v>0$, the results are\footnote{The analysis of the appendix did not establish the size of the error terms, and we have filled them in to match the mechanical values in Eqs.~\eqref{E12}--\eqref{N12}.}
\begin{align}
   E_{F\times} & = \frac{q_1 q_2}{v |t|}\frac{m_1^2 + m_2^2 + 2 m_1 m_2/\gamma}{E_0^2} + O(t^{-2}) \label{EFxresult} \\
    \bm{p}_{F\times} & = \frac{ q_1 q_2}{|t|}\frac{m_2^2-m_1^2}{E_0^2}\bm{\hat{z}} + O(t^{-2}) \label{pFxresult} \\
    \bm{L}_{F\times} & = O(t^{-2}) \label{LFxresult} \\
    \bm{N}_{F\times} & = \pm \frac{q_1 q_2}{\gamma^2v^2}\log\frac{m_2 +\gamma m_1}{m_1 + \gamma m_2}\bm{\hat{z}} + O(t^{-2}). \label{NFxresult}
\end{align}
Denoting these electromagnetic contributions with a subscript ``EM'', we see that the changes in electromagnetic conserved quantities (final minus initial) are 
\begin{align}
    \Delta E_{\rm EM} = 0, \quad \Delta \bm{p}_{\rm EM} = 0, \quad \Delta \bm{L}_{\rm EM} = 0,
\end{align}
and
\begin{align}
    \Delta \bm{N}_{\rm EM} = \frac{2 q_1q_2}{\gamma^2 v^2}\log\frac{m_2 + \gamma m_1}{m_1 + \gamma m_2} \bm{\hat{z}}. \label{DeltaNEM}
\end{align}
The total values of conserved quantities remain constant, but there is an exchange of mass moment between mechanical and electromagnetic degrees of freedom [Eqs.~\eqref{DeltaNmech} and \eqref{DeltaNEM}].

\subsection{Full time evolution}

Eqs.~\eqref{E12}--\eqref{N12} and \eqref{EFxresult}--\eqref{NFxresult} show that the initial and final values of the conserved quantities are
\begin{align}
    E = E_0, \quad \bm{p} = 0, \quad \bm{L} = - \mu b \gamma v \bm{\hat{y}}, \quad \bm{N} = 0. \label{allofthem}
\end{align}
Since there is no radiation in the problem (fields fall off like distance squared), it is clear that these values of the total (mechanical plus electromagnetic) conserved quantities must hold precisely at all times during the motion.\footnote{\label{foot:flux}Mathematically, one can easily show that there is no flux of energy, momentum, angular momentum, or mass moment through a large-radius sphere, because the relevant flux integrals fall off at least like inverse distance.}  It was therefore sufficient to calculate the initial values of the conserved quantities in order to know their values for all time.  Eq.~\eqref{allofthem} shows in particular that the conditions $\bm{p} = \bm{N} = 0$ defining the CEM frame [Eq.~\eqref{CEM-def}] indeed hold for our solution to the scattering problem.  

\begin{figure}
    \centering
    \includegraphics[scale=.65]{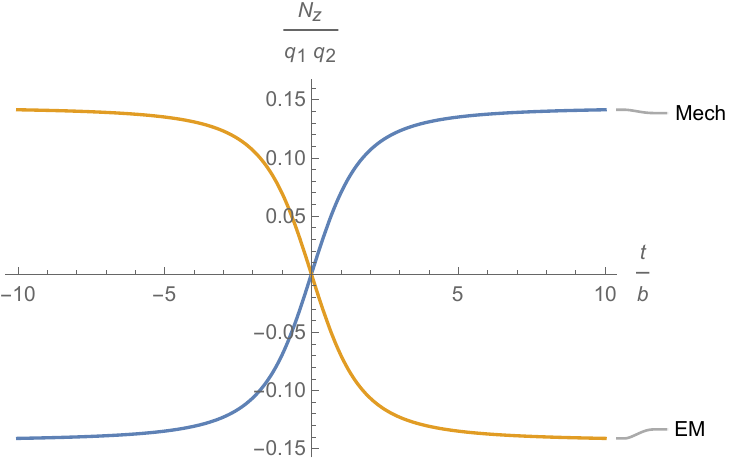}
    \caption{Exchange of mass moment between mechanical and field degrees of freedom in small-angle electromagnetic scattering.  We take speed $v=1/2$ and mass ratio $m_2/m_1=2$.}
    \label{fig:plot}
\end{figure}

We were unable to explicitly evaluate the electromagnetic contributions at intermediate times, since the problem is no longer effectively one-dimensional.  However, from the lack of radiation we know that the electromagnetic contributions are always equal and opposite to the mechanical ones,${}^{\ref{foot:flux}}$ which may be calculated from the trajectories \eqref{x1}--\eqref{z2}.  Fig.~\ref{fig:plot} shows the exchange in mass moment during the scattering process.

\subsection{Mechanical center of energy}\label{sec:CE}

The mass moment is somewhat unfamiliar, and a reader might question why we have considered it at all.  Given that the total momentum is zero, why not just consider the center of energy, which is an intuitive relativistic generalization of the center of mass?  The simplest answer---already given in the introduction---is that the mass moment is additive, so it makes sense to talk about separate mechanical and electromagnetic contributions.  However, one might still wonder about the behavior of the mechanical center of energy.  We will find that this behavior is quite misleading!

Let us define the mechanical center of energy as
\begin{align}
    \bm{C} = \frac{\sum_I E_I \bm{r}_I}{\sum_I E_I},\label{Cdef}
\end{align}
where as usual $I=1,2$ labels the particles.  Computing this quantity from our results [starting either with the trajectories \eqref{x1}--\eqref{z2} or the conserved quantities \eqref{E12}--\eqref{N12}], one finds
\begin{align}
    \lim_{t \to \pm \infty} \bm{C} = \frac{\mp q_1q_2}{\gamma^2 v^2 E_0}\left( \log\frac{m_2 + \gamma m_1}{m_1 + \gamma m_2}+\frac{m_2^2 -m_1^2}{E_0^2} \right) \bm{\hat{z}}.
\end{align}

The change in mechanical center of energy is thus
\begin{align}
    \Delta \bm{C} = -\frac{2 q_1q_2}{\gamma^2 v^2 E_0}\left( \log\frac{m_2 + \gamma m_1}{m_1 + \gamma m_2}+\frac{m_2^2 -m_1^2}{E_0^2} \right) \bm{\hat{z}}.\label{DeltaC}
\end{align}

This is a perfectly correct result given the definition \eqref{Cdef}, but it is hard to understand.   Because the center of energy is not a conserved quantity, there is no way to discuss exchange between particles and field.  And the specific form of \eqref{DeltaC} is quite puzzling because $\Delta \bm{C}$ does not vanish in the non-relativistic limit $v \ll 1$ (the second term survives).\footnote{To take the non-relativistic limit we must ensure that our small parameter $\chi_{\rm EM}$ \eqref{chiEM} remains small, which can be effected by expressing Eq.~\eqref{DeltaC} in terms of $\chi_{\rm EM}$ before letting $v \to 0$.}  It is well known that the mechanical center of mass is strictly conserved for non-relativistic two-body dynamics, where it is usually eliminated at the very start by passing to an effective single-particle description.  How, then, can the mechanical center of energy fail to be conserved in the non-relativistic limit?

The answer is that the mechanical center of energy \eqref{Cdef} does not actually reduce to the mechanical center of mass in the non-relativistic limit appropriate to the scattering problem.  No matter how small the velocity, there will be a correction to the particles' net kinetic energy that balances the potential energy $q_1 q_2/|\bm{r}_1-\bm{r}_2|$.  This correction falls off only like the inverse distance between the particles and hence  contributes to the mechanical center of energy (a distance-weighted average) even in the limit of infinite particle separation.  The problematic second term in \eqref{DeltaC} is precisely (twice) this contribution.  

We see that the mechanical center of energy does not have a very  useful non-relativistic limit.  By contrast, the mechanical mass moment properly reduces to the mechanical center of mass in the non-relativistic limit, in any frame with no net momentum.

\section{Discussion}

We conclude with some discussion of the character and implications of these results.  Let us begin with the size and direction of the electromagnetic scoot \eqref{NEM2}.  The dimensional scale of the effect is set by the charges and initial velocity in the combination $q_1 q_2/v^2$, which has units of mass moment (mass times distance).  The size also depends on the dimensionless Lorentz factor $\gamma$ and mass ratio $m_2/m_1$.  In particular, the masses influence the scoot only through their ratio, with the absolute mass scale playing no role in setting the size.  The direction of the scoot depends on the mass ratio as well as on the \textit{signs} of the charges, i.e., whether the interaction is attractive or repulsive.  If the interaction is attractive, the scoot is towards the final position of the heavier one, whereas if the interaction is repulsive, the scoot is towards the final position of the lighter one.

This association holds for the gravitational case \eqref{gravscoot} at sufficiently low velocities ($v < 1/\sqrt{3}$, to be precise), but breaks down above this threshold.  The qualitative agreement between the gravitational and electromagnetic results at small velocity is expected given the well-known gravitomagnetic analogy.  We have no intuition for the sign reversal in the gravitational scoot at $v=1/\sqrt{3}$, where the whole effect vanishes.  This reversal does not occur in electromagnetism.

In the introduction we commented that the gravitational and electromagnetic scoots have a universal character in that they are independent of impact parameter.   Our study of the electromagnetic case sheds considerable light on this issue.  The initial and final values of the mechanical mass moments are associated with logarithmic corrections to the particle position due to the long-range nature of the Coulomb force.  These corrections will appear at initial and final times irrespective of the details of the scattering encounter.  Similarly, our calculation of the electromagnetic field contributions relies only on the condition $|vt| \gg b$, which will occur at sufficiently early or late times in any scattering encounter.

In other words, our calculations show that \textit{whenever charged particles interact only by electromagnetic forces, there will always be non-zero mechanical and electromagnetic contributions to the mass moment, even in the limit of wide separation}.  These contributions are directed tangent to the particle separation, and hence will change in any scattering encounter that changes the orientation of the particles.  In particular, if $\bm{\hat{r}}_{12}^{\rm initial}$ is a unit vector pointing from particle $1$ to particle $2$ at early times and $\bm{\hat{r}}^{\rm final}_{12}$ is a unit vector pointing from particle $1$ to particle $2$ at late times, then there will be a change in CEM-frame mechanical mass moment given by\footnote{\textit{[Note added: If the particle Lorentz factors change as a result of the scattering, Eq.~\eqref{DeltaNgen} should instead read %assumes that for each particle, the initial and final Lorentz factors agree.  This assumption violated at sufficiently high perturbative order.  However, at those orders, there will still be a ``non-radiative scoot'' due to the difference between the initial and final  }
} \begin{align} \nonumber
\Delta \bm{N}_{\rm mech} = \left.\bm{\hat {r}}_{12} \frac{q_1q_2}{\gamma^2 v^2}\log\frac{m_2 + \gamma m_1}{m_1 + \gamma m_2} \right|^{\rm final}_{\rm initial}.\end{align}
}

\begin{align}
    \Delta \bm{N}_{\rm mech} = \frac{q_1q_2}{\gamma^2 v^2}\log\frac{m_2 + \gamma m_1}{m_1 + \gamma m_2} (\bm{\hat{r}}^{\rm final}_{12}-\bm{\hat{r}}_{12}^{\rm initial}).\label{DeltaNgen}
\end{align}
In small-angle scattering we have $\bm{\hat{r}}^{\rm final}_{12} \approx -\bm{\hat{r}}^{\rm initial}_{12}$, reproducing the result of our explicit calculation [Eq.~\eqref{NEM2} or \eqref{DeltaNmech}].  In general scattering (at higher order in perturbation theory, or without any approximation), there may be additional terms due to radiative losses, but the ``conservative'' contribution \eqref{DeltaNgen} will always be present.  In this sense the scoot is an unavoidable, and even trivial, consequence of the displacement between mechanical and electromagnetic mass moment that persists at large separation.

Why is this non-zero displacement present?  Here we can offer only mathematical reasoning.  The Coulomb force falls off as $1/d^2$ with increasing particle separation $d$.  This means that energies and momenta will receive corrections from the interaction at order $q_1q_2/d$, and the terms $E_I \bm{r}_I$ and $\bm{p}_I t$ present in the mass moment \eqref{Nmechintro} will have a finite limits at early and late times, where $d \sim \bm{r} \sim v t$.  Similar comments apply to the electromagnetic cross-term energy and momentum.  These various contributions to the total mass moment depend on a variety of parameters $(q_1,q_2,m_1,m_2,v)$, and it would be surprising if they were all individually zero at all parameter values.  We may set one linear combination to zero by choice of frame (the center of energy frame), but there will still be non-zero contributions from different degrees of freedom of the system.

These electromagnetic results provide context for the gravitational problem.  They provide encouragement that the gravitational result \eqref{gravscoot} is not an artifact of some peculiar choice of gauge but rather a bona-fide physical effect worthy of further exploration.  They also reveal that a proper accounting of the conserved quantities will undoubtedly require consideration of log corrections to particle position as well as gravitational field contributions.  These effects are surely an integral part of the initial and final configurations for the gravitational scattering problem, and will be relevant for any rigorous formulation of scattering as a map from past timelike infinity to future timelike infinity.   While the study of spacelike and null infinity in general relativity is rather mature, comparably little is known about timelike infinity, especially when matter is present.  We hope that our electromagnetic results will be helpful in establishing a  rigorous framework for the general relativistic scattering of massive bodies.  

\section*{Acknowledgements} We are grateful to Drew Milsom for helpful discussions.  This work was supported in part by NSF grant PHY--1752809 to the University of Arizona.

\appendix

\section{Field cross-term integrals at zero impact parameter}\label{sec:field-integrals}

In Sec.~\ref{sec:field-early} we showed that the leading behavior of the field cross-term integrals at early and late times may be determined from the integrals evaluated at zero impact parameter.  In this appendix we evaluate the relevant integrals.  We do not assume $v>0$ in this appendix.

Setting $b=0$ in Eqs.~\eqref{EI} and \eqref{BI} and changing to cylindrical coordinates $\rho^2=x^2+y^2$ and $\tan \phi=y/x$, we have
\begin{align}
\tilde{\mathbf{E}}_I & = \frac{q_I\gamma_I}{\tilde{R}_I^3}\left[ \rho \bm{\hat{\rho}} + (z-v_I t)\bm{\hat{z}} \right] \label{EItilde} \\ %(x-b_I, y, z-v_It) \\
\tilde{\mathbf{B}}_I & = \frac{q_I\gamma_I v_I}{\tilde{R}_I^3} \rho \bm{\hat{\phi}} \label{BItilde} \\
\tilde{R}_I & = \sqrt{\rho^2+\gamma_I^2(z-v_I t)^2},
\end{align}
where the tilde stands for evaluation at $b=0$.  The cross-term energy density is
\begin{align}
    \tilde{\mathcal{E}}_\times & = \frac{1}{4\pi} \left( \tilde{\bm{E}}_1 \cdot \tilde{\bm{E}}_2 + \tilde{\bm{B}}_1\cdot \tilde{\bm{B}}_2 \right) \\
    & = \frac{1}{4\pi} \frac{q_1 q_2 \gamma_1 \gamma_2}{\tilde{R}_1^3 \tilde{R}_2^3} \left( \rho^2(1+v_1v_2) + k_1 k_2 \right),
\end{align}
where we define
\begin{align}
    k_I = z - v_I t.
\end{align}
To compute the total energy we first perform the integration over $\rho$ and $\phi$, yielding
\begin{align}
    \tilde{E}_{F \times} & = \int \tilde{\mathcal{E}}_\times \rho d\rho d\phi dz \\
    & = \frac{q_1 q_2}{2}\int_{-\infty}^\infty  \frac{s_1 s_2+\gamma_1 \gamma_2(1+v_1 v_2)}{(s_1 \gamma_1 k_1  + s_2 \gamma_2 k_2)^2} dz,
\end{align}

where $s_I=k_I/|k_I|$ denotes the sign of $k_I$.  The remaining integrand is discontinuous at $k_1=0$ and $k_2=0$, and the integral must be split up at the corresponding values $z=v_1 t$ and $z=v_2 t$.  Performing these integrals, we find
\begin{align}
    \tilde{E}_{F \times} & = q_1q_2 \frac{1+v_1 v_2}{|v_1 t - v_2 t|} \nonumber \\
    & = \frac{q_1 q_2}{E_0^2 |v t|}\left( m_1^2 + m_2^2 + 2 \frac{m_1 m_2}{\gamma} \right). \label{EFxtilde}
\end{align}
In the non-relativistic limit $v \ll 1$, this reproduces the usual interaction energy $q_1 q_2/|z_1-z_2|$ of the one-dimensional problem.

The cross-term momentum density is given by
\begin{align}
    \tilde{\bm{S}}_\times & = \frac{1}{4\pi}\left( \tilde{\bm{E}}_1 \times \tilde{\bm{B}}_2 + \tilde{\bm{E}}_2 \times \tilde{\bm{B}}_1 \right) \\
    & = \frac{-1}{4\pi} \frac{q_1 q_2 \gamma_1 \gamma_2}{\tilde{R}_1^3 \tilde{R}_2^3} \left( (v_1+v_2)\rho^2 \bm{\hat{z}} - \rho(k_1v_2 + k_2v_1) \bm{\hat{\rho}} \right).
\end{align}
Following similar steps as before, the total momentum is 
\begin{align}
    \tilde{\bm{p}}_{F \times} & = \int \tilde{\bm{S}}_\times \rho d\rho d\phi dz \\
    & = - \frac{\gamma_1 \gamma_2 q_1 q_2}{2} \bm{\hat{z}} \int_{-\infty}^\infty \left( \frac{s_1 k_1 \gamma_1-s_2 k_2 \gamma_2}{\gamma_1^2k_1^2 - \gamma_2 k_2^2} \right)^2 dz \\
    & = q_1 q_2 \frac{v_1+v_2}{|v_1t-v_2t|} \bm{\hat{z}} \\
    & = \textrm{sign}(v) \frac{ q_1 q_2}{|t|}\frac{m_2^2-m_1^2}{E_0^2} \bm{\hat{z}},
\end{align}
where we note that $\textrm{sign}(v_1-v_2)=\textrm{sign}(v)$.

The angular momentum density $\bm{x} \times \tilde{\bm{S}}_\times$ is proportional to $\bm{\hat{\phi}}$ and has magnitude independent of $\phi$, so the total cross-term angular momentum vanishes,
\begin{align}
    \tilde{\bm{L}}_{F \times} & = 0.
\end{align}

The cross-term mass moment is built from the cross-term momentum, which we have already computed, together with the position-weighted average of the energy density.  The relevant integral for the latter is
\begin{align}
 & \int \tilde{\mathcal{E}}_\times \bm{x} d^3 x 
     = \frac{q_1 q_2}{2} \bm{\hat{z}}\int_{-\infty}^\infty  \frac{s_1 s_2+\gamma_1 \gamma_2(1+v_1 v_2)}{(s_1 \gamma_1 k_1  + s_2 \gamma_2 k_2)^2} z dz \nonumber \\
    & = q_1 q_2 \textrm{sign}(v_1t-v_2t)\bm{\hat{z}} \left(\frac{v_1+v_2}{v_1-v_2}+\frac{\log(\gamma_2/\gamma_1)}{\gamma_1^2 \gamma_2^2(v_1-v_2)^2}\right) \nonumber \\
    & = - q_1 q_2 \textrm{sign}(v t)\bm{\hat{z}}\Bigg(\frac{m_1^2-m_2^2}{E_0^2}+\frac{1}{\gamma^2v^2}\log\frac{m_1 +\gamma m_2}{m_2 + \gamma m_1}\Bigg). \label{int}
\end{align}

One subtlety of this calculation is worth noting. In computing the integral over $z$, we find that the anti-derivative is logarithmically divergent at $z \to \pm \infty$, with the divergence canceling out of the final answer.  This indicates that the integral over ``all space'' is not absolutely convergent and hence can depend on the manner in which the limit is taken.  By using cylindrical coordinates we have taken the limit using increasingly large cylindrical regions.  We have checked numerically that the answer is the same if we instead use increasingly large spherical regions.  We expect that any sufficiently symmetric choice of region will result in the same answer, and conclude that the cylindrical (or spherical) approach is a reasonable definition for the total electromagnetic mass moment in all of space.

The mass moment is given by adding $\tilde{\bm{p}}_{F\times}t$ to the integral computed in \eqref{int}.  This cancels the first term, leaving
\begin{align}
    \tilde{\bm{N}}_{F\times} = - q_1 q_2 \textrm{sign}(v t)\frac{1}{\gamma^2v^2}\log\frac{m_1 +\gamma m_2}{m_2 + \gamma m_1}\bm{\hat{z}}.
\end{align}
This completes the calculation of the cross-term field integrals at zero impact parameter. \vspace{-.5cm}

\bibliographystyle{utphys}
\bibliography{EMscootfinal.bib}

\end{document}